\documentstyle[11pt,epsf]{article}
\def\ba{\begin{eqnarray}\samepage}
\def\ea{\end{eqnarray}}
\input amssym.def
\input amssym.tex

\newcommand{\be}{\begin{equation}}
\newcommand{\ee}{\end{equation}}
\newcommand{\ben}{\begin{eqnarray}\displaystyle}
\newcommand{\een}{\end{eqnarray}}

\def\beq{\begin{equation}}
\def\eeq{\end{equation}}

\begin{document}
\title{Threebranes in twelve dimensions\footnote{Notes based on talk given at the  NATO
Advanced Study Institute conference on Strings, Branes and Dualities, Cargese, France, 26 May - 14
June 1997. }}
\author{Stephen F. Hewson\footnote{Email:
sfh10@damtp.cam.ac.uk}\\DAMTP, Silver Street, Cambridge, CB3 9EW\\ England}

\maketitle

\begin{abstract}
In this note we discuss   the theory of super-threebranes in
a spacetime of signature (10,2). Upon reduction, the threebrane
provides us with the classical representations of the M-2-brane and
the type IIB superstring.   Many features of the original super
(2+2)-brane theory
are clarified.  In particular,  the  spinors required to construct the
brane action 
and the (10,2) superspace are discussed.  
\end{abstract}

\section{Introduction}
Recently we have seen  great increases in the understanding of
non-perturbative string theory in terms of the eleven dimensional
M-theory  structure 
\cite{Duf:mtheory,Sch:dualitylectures}. This understanding has been
greatly enhanced by the realisation that many problems may be solved
by the introduction of a signature (10,2) `F-theory'
\cite{Vaf:ftheory}. To investigate this proposal further we discuss
the formulation of a Green-Schwarz  threebrane 
in twelve dimensions \cite{HewPer:22brane}.

\section{Supersymmetric threebrane in 12D}
Traditional supersymmetry $p$-branes propagate in a superspace which
is invariant under the action of some supersymmetric extension of the  
 Poincar\'{e} group. In 11D this essentially involves
the addition of the term
\begin{equation}\label{11dalgebra}
\{{Q_{11}^{\alpha}},{Q_{11}^{\beta}}\}=({\gamma}^p)^{\alpha\beta}P_p+
({\gamma}^{p_1p_2})^{\alpha\beta}Z_{p_1p_2}+
({\gamma}^{p_1\dots p_5})^{\alpha\beta}Z_{p_1\dots p_5}\,,
\end{equation}
where  $\gamma^p$ satisfy the 11D Clifford
algebra and $Q_{11}$ is a Majorana spinor. This algebra is maximal,
possessing (528+528) degrees of freedom, and also reproduces the type
IIA and IIB algebras  with the help of
dimensional reduction and T-duality \cite{Tow:fourlectures}. 
However, all of these algebras 
may be obtained via a
reduction of another maximal system: the `F-algebra', given by 
\begin{eqnarray}\label{algebra1}
\{Q^\alpha,Q^\beta\}_{(10,2)}&=&\frac{1}{2!}\left(C\Gamma^{\mu\nu}\right)^{\alpha\beta}Z_{\mu\nu}+\frac{1}{6!}\left(C\Gamma^{\mu\nu\rho\sigma\lambda\delta}\right)^{\alpha\beta}Z^+_{\mu\nu\rho\sigma\lambda\delta}\nonumber\\
\left[M_{\mu\nu},M_{\rho\sigma}\right]&=& 
M_{\nu\sigma}\eta_{\mu\rho}
+M_{\mu\rho}\eta_{\nu\sigma}-M_{\nu\rho}\eta_{\sigma\mu}-
M_{\sigma\mu}\eta_{\nu\rho} \,,
\end{eqnarray}
where $Z_{\mu\nu}$ and $Z^+_{\mu\mu\rho\sigma\lambda\delta}$ are central
charges,  $Z^+_{\mu\mu\rho\sigma\lambda\delta}$ being
self-dual, and  $Q^{\alpha}$ is Majorana-Weyl. This seems to be a natural
algebra to study in twelve 
dimensions \cite{Hew:ftheory}, and leads to the existence of a
threebrane solution. We initially formulate the superspace underlying
the F-algebra and then discuss the formulation of the threebrane\footnote{In the following discussion of the properties of
the threebrane we set the central  6-form charge to zero for
simplicity.}.  

\subsection{The Superspace}
Given a super Lie group $G$ and some subgroup $H$ we may define a
coset 
supermanifold ${\cal{M}}=G/H$. Ordinary superspace is formed by
considering the coset of the super Poincar\'{e} group by the Lorentz
rotations. In 12D we perform an analogous procedure for
the F-algebra (\ref{algebra1}). We define 
\begin{equation}
G(X,\theta,\omega)=\exp(\theta_\alpha
Q^{\alpha}+X^{\mu\nu}Z_{\mu\nu})\exp(\frac{1}{2}\omega^{\mu\nu}M_{\mu\nu})\,.
\end{equation}
To generate an infinitesimal change in the coordinates we act on the
left with an infinitesimal group element $G(\delta
X,\delta\theta,\delta\omega)$. From these changes we may generate
invariant forms on the superspace
\begin{equation}\label{forms}
\Pi^\alpha=d\theta^\alpha\,,\quad\Pi^{\mu\nu}=dX^{\mu\nu}-\frac{1}{2}\theta_\alpha(\Gamma^{\mu\nu})^{\alpha\beta}d\theta_\beta\,.
\end{equation}
It is natural to consider the F-algebra as descending
from a curved space de Sitter algebra \cite{HolPro:}, in which the 2-form
$Z_{\mu\nu}$ obeys the same algebraic relationships as the Lorentz
generator $M_{\mu\nu}$ with itself and $Q^\alpha$. Due to the
non-commuting nature of the 
$Z_{\mu\nu}$ generator in this case, the action of the supergroup on
the manifold is 
highly non-linearly realised. However, a careful analysis of the
expansion of the transformed group elements shows that we may obtain a
{\it linear} realisation of the supersymmetry if one of the following
conditions is met:
\begin{equation}\label{projection}
\mbox{(i)}\,\,\,(\Gamma_{\mu\nu})^{\alpha{[}\beta}(\Gamma^{\mu\nu})^{\gamma{]}\delta}=0\,\hskip2cm
\mbox{(ii)}\,\,\,{\widetilde{{\cal{P}}}}C\Gamma^{\mu\nu}{\cal{P}}=0\,.
\end{equation}
In  case (i) the invariant forms are still written as in
(\ref{forms}), whereas if the  so-called `purity' condition (ii) is
met then the spin space becomes  trivial:
\begin{equation}\label{forms2}
\Pi^\alpha\rightarrow d\theta^\alpha\,,\quad\Pi^{\mu\nu}\rightarrow
dX^{\mu\nu}\,. 
\end{equation}

\subsection{The threebrane}
The basic threebrane action in the new superspace is  defined in exact
analogy with 
the traditional Green-Schwarz string and membrane:
\begin{equation}\label{action}
S=\int
d^{4}\xi\sqrt{\det({\Pi^{\mu\nu}}_i{\Pi_{\mu\nu i}})}\,,
\end{equation}
where the ${\Pi^{\mu\nu}}_i$ are the pullback of the forms
(\ref{forms}) to the (2+2)-dimensional worldvolume. 
 By construction,
this algebra is manifestly {\it spacetime} supersymmetric, whereas in order
for {\it worldvolume} supersymmetry to occur we must obtain a matching of
bosonic and fermionic excitations on the brane.  To see that this is
the case even for the F-algebra, note that a superspace consists of
the disjoint union of a bosonic piece ${\cal{B}}$ and a fermionic
piece ${\cal{F}}$.  The
supersymmetry generator $Q$ maps ${\cal{B}}$ onto a proper subspace of
${\cal{F}}$ and vice-versa. Due to the group structure of the
superspace this implies that the mapping is a bijection, and therefore
the bosonic and fermionic degrees of freedom must match up. The
(2+2)(=3)-brane  has eight transverse  scalar fields
propagating on 
the worldsheet, whereas the on-shell Majorana-Weyl spinors have
sixteen real degrees of freedom. In order to obtain a match we must
project out half of the remaining spin degrees of freedom. For
super-Poincar\'{e} brane theories, the 
additional fermionic excitations  are lost due to a
$\kappa$-symmetry of the full action consisting of a Dirac term of
type (\ref{action}) plus a Wess-Zumion integral
\cite{Tow:fourlectures}. No such symmetry 
 may occur in the 12D theory \cite{HewPer:22brane}. However, to obtain a linear
realisation of the superspace theory of the underlying {\it de
Sitter} algebra we need to make a projection of the spinor operators
such that (\ref{projection}) holds. This has the same effect as a
$\kappa$-symmetry and gives us the required (8+8) worldvolume degrees
of freedom. Note that since branes are microscopic objects, and
therefore insensitive to large scale structure, we choose to keep this
restriction in the flat space limit.

\subsubsection{Reduction to lower dimensions}
We now consider the reduction of the 12D threebrane. We reduce on the
timelike direction 12 to 11D and on a null torus to  10D. Reduction to
10D and 11D 
requires us to act on the spinor indices with the projectors
${\cal{P}}_{10}=\frac{1}{2}(1+\Gamma_0\dots\Gamma_9)$ and
${\cal{P}}_{11}=\frac{1}{2}(1+\Gamma_0\dots\Gamma_9\Gamma_{11})$
respectively. It is a simple matter to show that one obtains  two 10D
spinors of the {\it same} chirality and one 11D spinor after these
reductions.  Hence the theories are projected down onto the IIB
 and 11D sectors of M-theory. We now consider the reduction of the
action (\ref{action}), concentrating on the circle compactification to
11D.  Upon
reduction we find that the $Z^{p12}\rightarrow P^p_{11}$ and
$Z^{pq}\rightarrow Z^{pq}_{11}$, in an obvious notation. 
There are two ways in which the superspace forms will reduce. In the
case for which the first equation of (\ref{projection}) holds then the
forms possess a torsion term as in (\ref{forms}) and reduce as follows
\begin{equation}
\Pi^{p12}=dX^{p12}-\frac{1}{2}\theta_{\alpha}(\Gamma^{p12})^{\alpha\beta}
d\theta_{\beta}\longrightarrow\Pi^{p}=dX^{p}-\frac{1}{2}\theta_{\alpha}(\gamma^{p})^{\alpha\beta}
d\theta_{\beta}\,,
\end{equation}
where we set the superspace terms $Z^{pq}_{11}$ equal to zero as
usual. Thus we obtain the usual Green-Schwarz actions in lower
dimensions. In the case for which the spinors are restricted so as to
satisfy the equation
${\widetilde{{\cal{P}}}}C\Gamma^{\mu\nu}{\cal{P}}=0$  the 
superspace is trivially realised; there then exists a
very simple (2+2)-brane action  which possesses a manifest supersymmetry
between the bosons and fermions:
\begin{equation}
S=\int d^4\xi \sqrt{\det(\Pi^A_i\Pi^B_j{\cal{G}}_{AB})}\,,
\end{equation}
in which the superspace `metric' is defined by the union of the flat
spacetime metric $\eta$ and the charge conjugation matrix $C$, as
${\cal{G}}=\mbox{diag}(\eta,C)$. The forms in the action are now trivial:
$\Pi^A=(dX^\mu,d\theta^\alpha)$.  In this situation it is less clear
that the action will reduce to the correct {\it super-Poincar\'{e}}
invariant action under dimensional reduction. However, when one
performs a superspace compactification then it becomes clear that a
natural torsion term of the form
$\theta_{\alpha}(\gamma^\mu)^{\alpha\beta} d\theta_{\beta}$
may be  introduced into the expression for  $\Pi^\mu$
\cite{HewPer:22brane}. Thus we do 
still recover the 
traditional Green-Schwarz  actions in lower dimension. However, the
addition of this torsion term is not compulsory and thus reduction may
also lead to a  torsion-free superspace sector of M-theory. A
discussion of such supersymmetries is presented elsewhere
\cite{Hew:generalised}.

\section{Conclusion}
We have outlined the formulation of the  three-brane in twelve
dimensions. Upon reduction we find that the superspaces associated with
the F-algebra 
reduce to those for the Poincar\'{e} algebra,  and that the
threebrane action reduces to that for a usual Green-Schwarz string or
membrane. In addition,  
reduction of the 12D spinors leads to one spinor in 11D and two
spinors of the {\it same} chirality in 10D.  Thus we obtain the
classical representations of the   type IIB string and the
M-2-brane, directly from twelve dimensions.

\section{Acknowledgements}
I would like to acknowledge financial support from NATO and Queens'
College, Cambridge. I also extend my sincere thanks to the organisers
of the 
Advanced Study Institute conference  on Strings, Branes and Dualities, Cargese, France, 26 May - 14
June 1997, where this work was presented. Finally, I would like to
thank Dr. Maria Woodfield, for useful discussions.

\end{document}